# Defining and Adopting an EUC Policy: A Case Study

Roger Turner

Wesleyan Assurance Society, Colmore Circus, Birmingham, B4 6AR
roger.turner@wesleyan.co.uk

**ABSTRACT**

*End User Computing (EUC) carries significant risks if not well controlled. This paper is a case study of the introduction of an updated EUC policy at the Wesleyan Assurance Society. The paper outlines the plan and identifies various challenges. The paper explains how these challenges were overcome.*

*We wrote an EUC Risk Assessment Application which calculates a risk rating band based upon the Complexity, Materiality and Control (or lack of it) pertaining to any given application and the basis of assessment is given in this paper.*

*We find that EUC applications are clustered in certain business areas and this information supports the need for addressing these risks on a wider scale with a view to improving overall business efficiency.*

*The policy uses a risk-based approach for assessing and mitigating against the highest risks first and obtaining the quickest benefit.*

*A Business As Usual (BAU) process has been put in place to monitor activity and we are seeing an improvement in the quality of EUC in the Society.*

## 1 INTRODUCTION

The paper gives the background and the case history of the introduction of an EUC policy at the Wesleyan Assurance Society.

The Wesleyan Assurance Society is a UK based financial services mutual founded in 1841 that provides specialist advice and solutions to doctors, dentists, teachers and lawyers. Wesleyan aims to build life-long relations with its customers, providing them with products and services at every stage of their life from graduation to retirement and beyond.

The Wesleyan group of companies employs approximately 1,500 staff divided between the Head Office in Birmingham, Oswestry, New Malden and Northwich as well as sales staff located throughout the UK.

There are £7.6 bn assets under management and the Society is successful in passing on good performance to its policyholders through its financial strength and long-term investment policy.

The EUC policy covers any application not supported by IT and 90% are spreadsheets. In this paper we refer to them all as "applications" for complete coverage.

After the Action plan there are the two main sections. These detail the challenges and how they are overcome, then an EUC Risk Assessment Application specially written as part of the policy is described in section 10. At the end of this paper there is a graphic displaying some research which was done at the outset of the project.

## 2 BACKGROUND

Before the 1980's all serious computing was done under the control of the organisation's IT department where it was best practice for strict controls to be in place for the design, development and maintenance of all the organisation's systems and programs.

Any new systems or changes to systems which were required by the business would frequently be done according to a lengthy development life cycle. Sometimes the requirements would change while the system was being developed so that the new system was not what the customer wanted. Business units doing their own thing was not an option.

Microsoft Excel first became available in 1985 (Wikipedia, 2018) and its gradually increasing functionality and use provided opportunity for computing independently of the organisation's IT department to take place. EUC was born along with its associated risks.

By the end of 2017 the EUC policy had been successful in two pilot runs and had been reviewed by the Chief Operating Officer (COO) as sponsor. He approved the policy being rolled out starting Q2 2018.

EUC control is part of the Data Governance function at Wesleyan.

## 3 CASE HISTORY AT THE WESLEYAN – ASSOCIATED CHALLENGES

The Wesleyan Assurance Society updated its EUC policy in 2012 because of Solvency II. Subsequently the Data Governance department reviewed the effectiveness of the policy and determined that changes were required to enhance its use.

The Society had not experienced any problems with its spreadsheet and end user applications but was keen to ensure it enhanced its policy to keep pace with best practice and minimise the risk of issues arising in the future.

The business need which brought about the update of the policy was the potential risk of an application causing a substantial loss event. We considered this to be enough to warrant at least an investigation into how best to mitigate this risk.

The objective is to establish a clear plan of action, try it using at least one pilot and, once proven, roll it out throughout the Society. Two pilot runs were successful and the Society approved a phased roll-out on risk-based approach.

The main challenges which were faced were:

a. What to cover in the Policy (the Scope)

b. Where the EUC risks are (Identifying the business areas at greatest EUC risk)

c. Obtaining buy-in from each business area to adhere to and implement the EUC policy

d. Establishing effective storage and retrieval of EUC application metadata

e. Getting managers to record applications which fall short on Magique (the Society's already existing risk management system) (Magique, 2018)

f. Regular review.

These challenges are not untypical across the industry. An example of this is the experiences faced by Chambers and Hamill in respect of understanding where the EUC risks are in a banking environment and assigning responsibility for them (Chambers & Hamill, 2008).

## 4 PLAN OF ACTION

We considered the proceedings of a Hellenic American Union conference (Mallikourtis & Papanikolaou, 2010) and attended a workshop run by The Corporate IT Forum (CITF, 2016). Resulting from this background we decided that the following steps should be taken:

a. Produce the first draft of the updated Policy Document which includes a means of assessing applications (spreadsheets) for risk. The scope to cover "any computing which is not supplied by, acquired by or supported by any of the Wesleyan's formal IT departments". As to the applications, approximately 95% the EUC applications are spreadsheets. The scope is not complete, however, without including local databases (usually Access), Business Intelligence reports (e.g. SQL, Crystal, Power BI), Mobile apps and some third-party apps.

b. Find stakeholders who are willing to co-operate in running at least one pilot.

c. Run the pilot(s) which involves collecting data about each submitted EUC application, assessing it for risk and storing the details in a repository where it could be accessed when the need arises.

d. Conduct "show and tell" sessions to demonstrate which applications already have satisfactory controls and which might be deemed to fall short.

e. Agree an action plan to fix any errant applications.

f. Apply governance which will then become part of the EUC policy.

Several challenges and how they were overcome are documented here.

## 5 CHALLENGES FACING EUC ROLL-OUT & OVERCOMING THEM

### 5.1 Defining the Risk Metrics to assess the applications with

The Complexity and the Materiality of an application are the two main contributors to risk. The more complex a spreadsheet (or for that matter any application) is, the greater the risk is of the risk crystallising and creating an issue. Once the risk crystallises, how material is the effect on the Society's business operation?

## Complexity

We used one of the simpler ways to measure Complexity and this is suggested by PwC (PwC, 2004). A spreadsheet with low complexity is just for information logging and tracking. There are no formulae or links. Medium complexity is where simple formulae are used, for example to translate or reformat information. High complexity is the rest, where complex formulae are used, there are links to external sources, macros and modelling.

The more complex an application is the less likely someone other than the author can understand it and the greater is the spreadsheet risk.

## Materiality

Materiality could be measured as the impact resulting from the risk crystallising. This could be:

a. Inconvenient

b. Poor Customer Outcomes

c. Reputational

d. Loss of Business

e. Financial

f. Statutory / Legislative

Different areas of the business rank these in different orders so we used a different approach instead.

Independent research done by Chartis suggests the following classification for materiality (Chartis, 2016):

a. High – Application supports financial or regulatory reporting or private or confidential information.

b. Medium – Application supports management reporting, calculation or input into a core management information system, or used for making key business decisions.

c. Low – internal operations or day to day decisions, or contains outputs from core management information systems.

## Control

Following the Complexity and Materiality metrics in this way leads us to the front face of the cube provided that the application is well controlled.

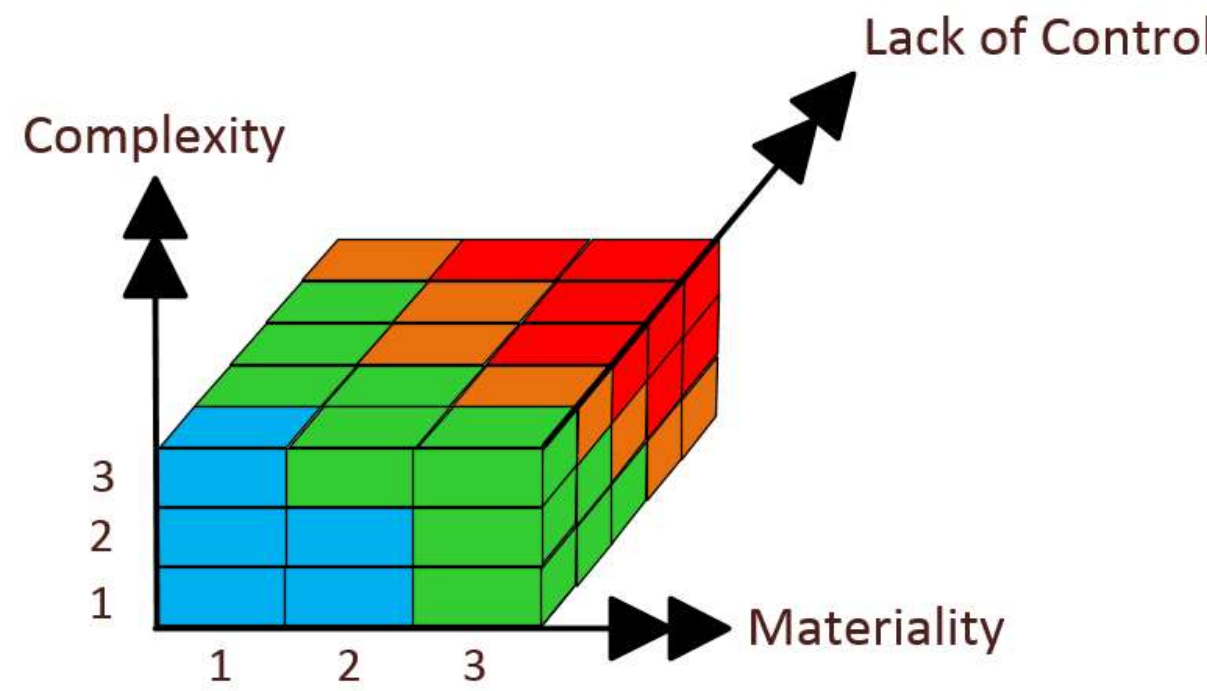

Figure 1- Complexity, Materiality & Control Metrics

The four colours on the cube indicate the risk rating band, and this identifies what remedial action, if any, is needed ranging from blue (none) to red (urgent action needed). Wesleyan uses the Magique system to record the risks and track them through to a resolution. (Magique, 2018).

Control, when considered, defines how far back we go in the cube.

We wrote a Risk Assessment Application for the EUC policy. The SMEs or other experts in the user department are provided with the application (which itself is a spreadsheet) then they use the application to assess the risk. Complexity and Materiality of an application are collected by the SMEs' self-assessment because they have hands-on knowledge of the applications and their context within the business.

By a series of yes / no questions the risk assessment application then gathers information about:

a. How accessible the application is, whether its location is known and whether there are operating instructions

b. Business Continuity, Back-up and Recovery

c. Version controlling, whether it needs reviewing and evidence of having been tested

d. Security, Privacy and Integrity, in other words unauthorised access to the system

e. The ability to fix the application if it breaks, including the existence of a second person able to fix and the existence of technical documentation

f. Finally, whether the system contains personal or sensitive personal information (in the context of GDPR, the General Data Protection Regulation, (IT Governance, 2018)).

The answers to the questions are recorded in the Risk Assessment Application and the application calculates the risk rating band. The user then sends the result back to Data Governance who records the results and ensures that there is an action plan to fix the application if it falls short within the assessment.

### 5.2 Whether to use a Top Down or Bottom up approach

As regards knowing what to assess for risk, two approaches are available, one being top down and the other being the bottom up. The bottom up approach means scanning the whole of the file store for spreadsheets, databases and such like for likely candidates and then finding owners. Even though there are tools which can scan for spreadsheets (Microsoft (2013), Finsbury (2014)) this is a formidable task if one considers that there could be several million files, only a few of these in current use and a few again requiring assessment.

The other way is to use the top down approach where managers and subject matter experts know where their applications are and can use the Risk Assessment Application to assess their applications and return the results. This is what we believe to be a more practical method.

### 5.3 Assimilation of the EUC policy

The full version of the policy document came to more than 80 pages and reading this is a big ask. We considered that effective communication of the policy is important so we split the document into smaller, more manageable amounts and put these on the intranet to draw the reader's attention to what action is needed based on their role, being one of the following:

a. Executives

b. Senior Managers

c. Managers

d. Subject Matter Experts (SMEs)

e. Data Stewards

For example, if the reader is a manager, the manager is led to the screen in Table 1.

Instructions are on the left and the hyperlinks on the right reference the appropriate part of the policy. The hyperlink for "Risk Assessment App" launches the application in Excel and they save it so that the users can use this risk assessment application to assess their applications.

## END USER COMPUTING ACTION REQUIRED OF MANAGERS

| Action Required | Resource |
|---|---|
| Adhere to the End User Computing policy and the Data Governance Policy, Cascade to your direct reports. | EUC Policy<br>Data Governance Strategy |
| Understand the Risks and Controls (Magique) for End User Computing | Risks<br>Controls |
| Understand the impact of end user computing on data quality. | Data Governance Strategy |
| Responsible for Data Quality including fitness for purpose of End User Computing applications, and making these known to Data Governance. Responsible for improving the control. | Taking Remedial Action |
| Provide assurance to your Senior Manager / Executive through an assessment of 'data' risk expressed within your Operational Risk and Control Assessments (ORCAs); or, where assurance cannot be given provide an action plan to strengthen controls/contain and fix known issues. | |
| Identify all end user computing applications so that the inventory of all End User Computing applications kept by the Senior Manager is kept up to date. Review each application periodically as required by the next review date. | EUCA Inventory |
| Ensure that each end user computing application is assessed to determine its risk rating and that testing has been done. Communicate the results to Data Governance. | Risk Assessment App.<br>Operating Instructions<br>Risk Metrics |
| Assist in creating and implementing an action plan to fix errant applications as indicated by the assessment tool. | Action Plan |
| The Complexity and Materiality of each EUC application are each graded 1 to 3 and if the sum of these is 5 or more then the Development Life Cycle must be understood and followed as appropriate. | Development Life Cycle |
| Establish a control framework which ensures the above actions are met. | |

Table 1 Sample Intranet page - EUC action required of managers

### 5.4 How to engage the Stakeholders

It was known that complex spreadsheets can contain many errors (Bregar, 2004) and the challenge was to maintain the buy-in from the stakeholders to mitigate against potential spreadsheet risk.

We decided to run pilots with two willing departments, chosen for the likelihood of having material or complex applications. Both of these were in the Finance area, one being Middle Office and the other Financial Accounting so we had to approach the department heads for their cooperation.

It certainly helped to have a well-prepared presentation identifying the risks and benefits surrounding EUC applications.

To facilitate buy-in we made and used a "horror slide" to highlight the risks in which some firms have lost billions of dollars, because of a mistake in a spreadsheet. Ample evidence is found on the (EuSpRIG, 2018) page.

We pointed out that writing a spreadsheet can be a quick and easy solution but the costs in the event of a risk crystallising can be substantial.

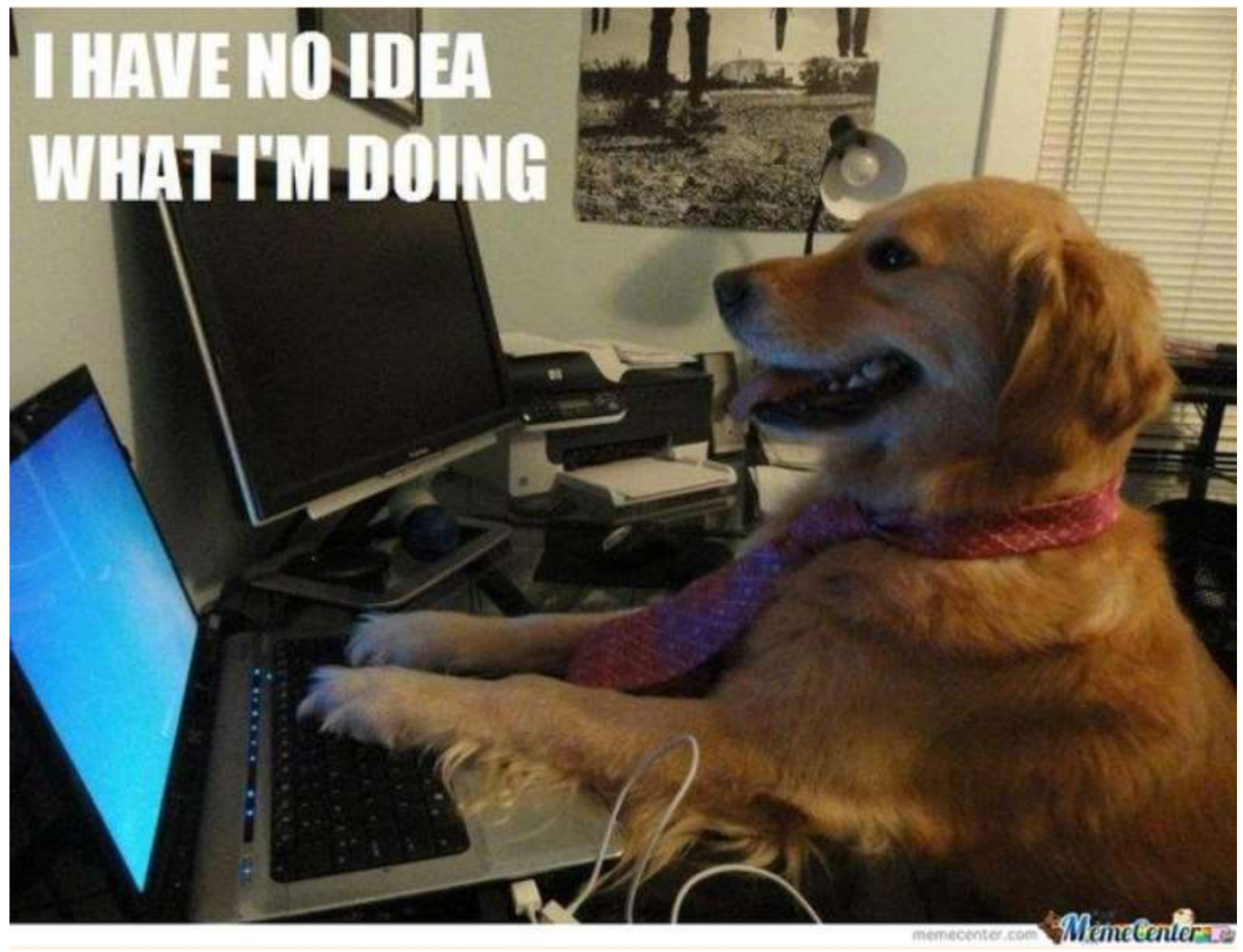

Figure 2- Engaging willing but unknowledgeable users

Think of a situation where only one person knows how to run an application, and that person is not there when the result is needed. The temptation is to get anybody to do the job, not knowing what to do or how to do it. We made the point by using this picture from the (Financial Times, 2013) showing this willing but unknowledgeable user.

Both pilots ran for five weeks during which the departments had each submitted 20 applications. We gave them the Risk Assessment application (see section 10 for the screen shots) and we collected the details of all the applications submitted (they were all spreadsheets). The collection for each application took around 10 minutes.

### 5.5 Assessment Results Returned to Data Governance

Each application returned had a risk rating calculated from details provided by the SMEs and the opportunity was available to challenge some detail if thought needed. For the complexity metric several tools are available.

Excel Inquire (Microsoft, 2016) is the easiest mentioned here and can report on links between spreadsheets and worksheets, and identify errors, hidden sheets and such like.

### 5.6 The Show and Tell sessions

Even the users' own assessment of the applications in the pilots gave a surprising proportion of applications with a red rating meaning that urgent action is needed to fix to mitigate potential risk. We looked at the reason why a poor rating was being produced and if there were any quick wins to remediate. The applications concerned all had either the materiality or complexity set to 3 with the other parameter at least 2. In these applications, the main concern was expressed in the security section where the functions or data in the spreadsheets could be open to accidental alteration or corruption (although we found no evidence this has actually occurred), and in some cases there was lack of version control.

Fixing these was seen as a quick win because spreadsheets could be baselined and copies made read only and before the next use a comparison could be made with the baseline. Comparison against a baseline can be done using Excel Inquire (Microsoft, 2016).

After the quick wins several red applications became amber and the most frequent reason for them remaining amber was the lack of evidence for testing and sometimes the lack of technical documentation (as opposed to any ongoing concerns).

This is more of an ongoing issue, however, Finance asserted that the results of these applications are subject to audit and there are many professionals who are equipped to challenge the results should any be suspicious.

The outcome is that the Risk Assessment Application highlights areas where attention to the control of an application ought to be focussed and it is up to the user department as to what action to take. They are responsible for a truthful entry of data into the Risk Assessment application and are accountable for whatever risks there are in the EUC applications.

### 5.7 Where and how to store the assessment results of EUC applications

Wesleyan's Group Reference Architecture provides for the use of Orbus' iServer as a repository for all the assets, whether an IT system or part of EUC. (Orbus, 2018).

Each application, (spreadsheet, other EUC application or IT supported system) can be stored in a way whereby its relationships with others can be visualised, for example in terms of the processes the application is used by, which department runs the process and which technology or platform the application runs on.

Its use within EUC is to be able to report on applications which require remedial action and to trigger action when an application needs to be reviewed. The policy states that each application should be reviewed annually.

We defined the data dictionary for EUCAs and arranged for iServer to have corresponding attributes for this metadata. It made sense to make iServer the master repository for EUCA information.

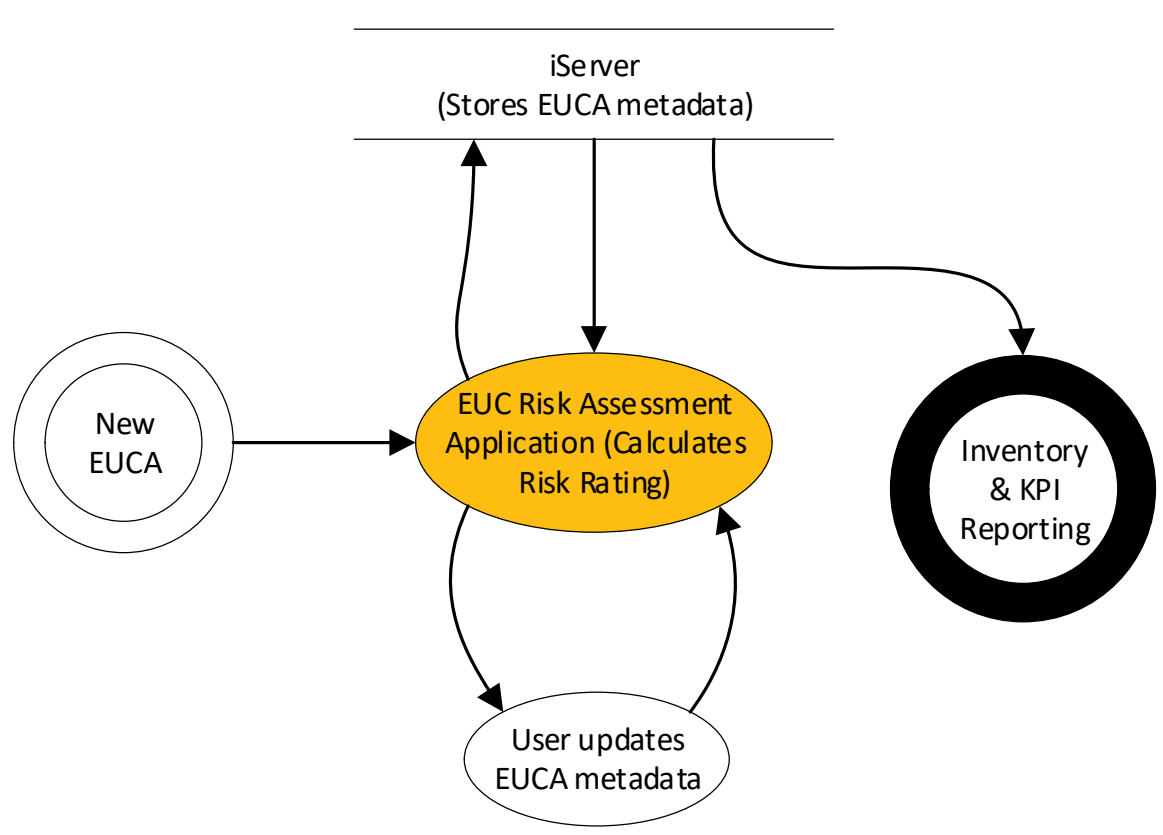

Figure 3: EUCA Metadata Flow

The user records one or more EUCAs in the Risk Assessment Application which calculates the risk rating band and advises if any remedial action is needed and why. The user passes the completed Risk Assessment Application to Data Governance who uploads the metadata into iServer. iServer exports metadata about a selection of or all the EUCAs for the inventory, KPI reporting and re-assessment by the Risk Assessment Application. The user can be given a copy of the Risk Assessment Application containing his or her own data for updating and returning to Data Governance for upload again into iServer.

## 6 ROLLING OUT THE EUC POLICY

We rolled out the EUC Policy in several stages.

### 6.1 Head Office Managers Meeting, April 2018

This monthly meeting at the Wesleyan consists of approximately 200 managers and the COO had given permission for us to deliver a short presentation on EUC at that meeting. The presentation focussed the audience on the risks posed by EUC (Financial Times, 2013) and by citing some of the public loss events (Chartis, 2016). Then the action was given to everyone to

- Read the EUC policy (which was already on the intranet)
- Complete a short assessment template by 30 June 2018 which was issued to everyone immediately after the meeting (See screenshot in Appendix B)
- Flag any high-risk applications to Data Governance as soon as possible.

The assessment template is an Excel spreadsheet which is applicable to the whole of the manager's department. The main point is that it is easy to complete and it asks for the manager to identify the process which has the most complex calculations or which has the most material impact on the business.

The spreadsheet asked the managers to assess that process according to the metrics already established. Then they had to return the completed template to Data Governance by 30 June 2018. Very often the managers enlisted the help of their Subject Matter Experts and other staff to select and assess their most complex processes. The spreadsheet provided the manager with one of three possible messages:

- You are Green. Please return this spreadsheet to Data Governance - no further action needed, however you are accountable for the results which you have returned. Any incidents as a direct result of spreadsheet errors that impact on a material process will need to be reported to Data Governance urgently.
- You are Amber. Action is needed. Return this spreadsheet to Data Governance. Your spreadsheets and applications need to be assessed, errant ones recorded on Magique and there needs to be an action plan to fix.
- You are Red. Urgent action is needed. Return this spreadsheet to Data Governance. Your spreadsheets and applications need to be assessed, errant ones recorded on Magique and there needs to be an urgent action plan to fix.

Figure 2 (right) shows the number of assessments returned in each category. Forty-six of the replies were to the effect that the department had no EUCAs to assess. There were an additional 69 replies (not included in Figure 2) saying that their area was within someone else's and to include it would be a duplication.

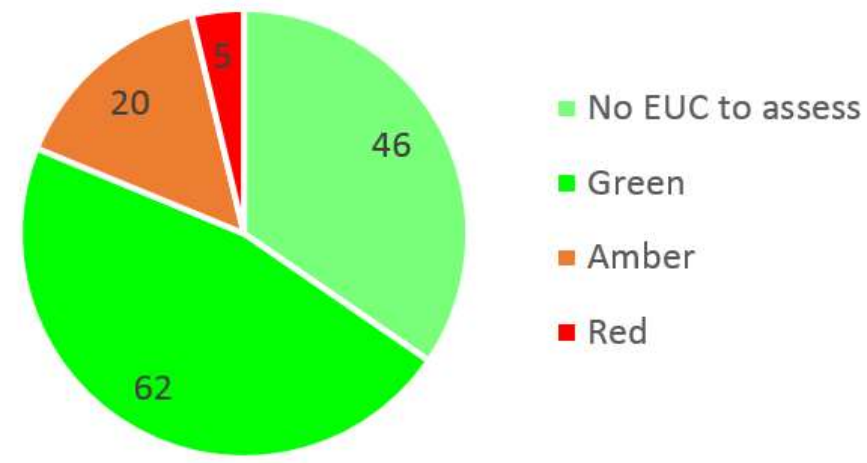

Figure 4: Assessment Returns by Department

### 6.2 Collating the returns and publicity, August 2018

By the end of July all the results were in (not without a certain amount of gentle persuasion including walking the building and meeting the stragglers face to face)! This was the time for publicising the importance of controls on EUC and having an agenda item on various Data Governance meetings to update people on the progress. Publicity was helped by the author of this paper being shortlisted for a CIR Risk Management award in the Newcomer of the Year category (CIR Magazine, 2018).

The next important deadline was to table whatever remedial actions to be taken on applications falling short of the controls at the Group Executive meeting in November 2018 so that remedial action could be put into each departments' 2019 business plans.

### 6.3 Drilling down to expose further risks, September and October 2018

We identified those departments which had an assessment of red or amber and issued them with the full EUC Risk Assessment Application to identify any applications which fall into the amber or red category, therefore requiring remediation.

Chambers & Hamill sets out certain minimum controls (Chambers & Hamill, 2008). We found that an important control is the ability to support an application once the author has ceased to be available to provide that support, especially without leaving technical documentation which the new incumbent would find virtually essential.

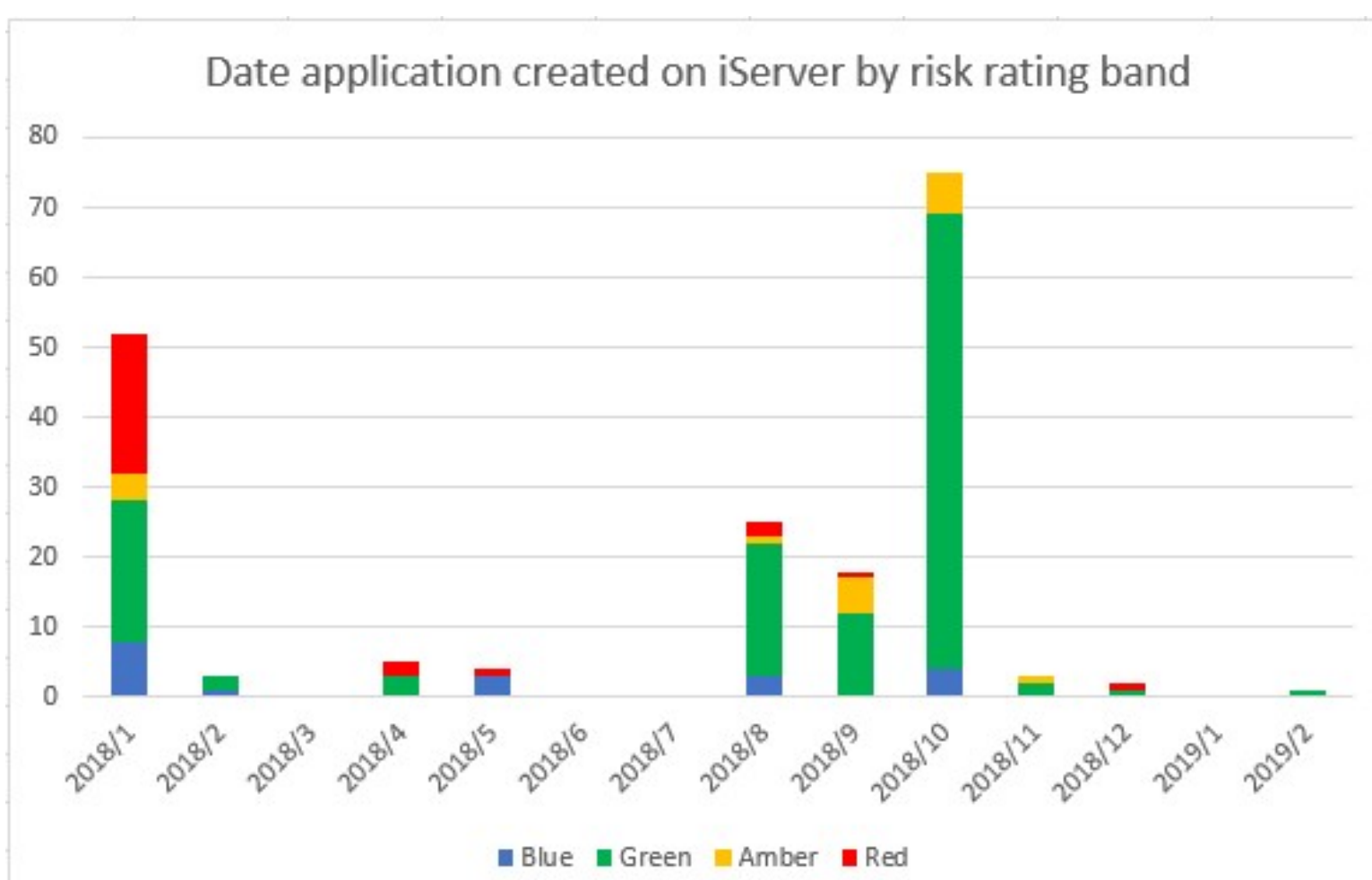

Figure 5: Applications Recorded on iServer

Figure 3 shows when EUC applications were created on iServer. All the assessment templates were completed by department heads or their representatives between May and July 2018 so those entered before this time had resulted from the pilots and other sources. The presentation to the Executive (next section) covered everything on iServer up to October apart from the pilot in Financial Accounting and Solvency Reporting in recognition of the fact that they have separate controls.

During this period, we gave priority to ensure that iServer was kept up to date with all changes to do with EUC applications and iServer became the master repository for this data. The schematic in Figure 3 shows the flow of data between the Risk Assessment Application, iServer and elsewhere.

### 6.4 Presentation to the Executive, November 2018

A short paper was presented at the Group Executive Meeting.

The paper included Figure 6, a table of risk rating v impact relating to the 158 EUC applications extant at the time.

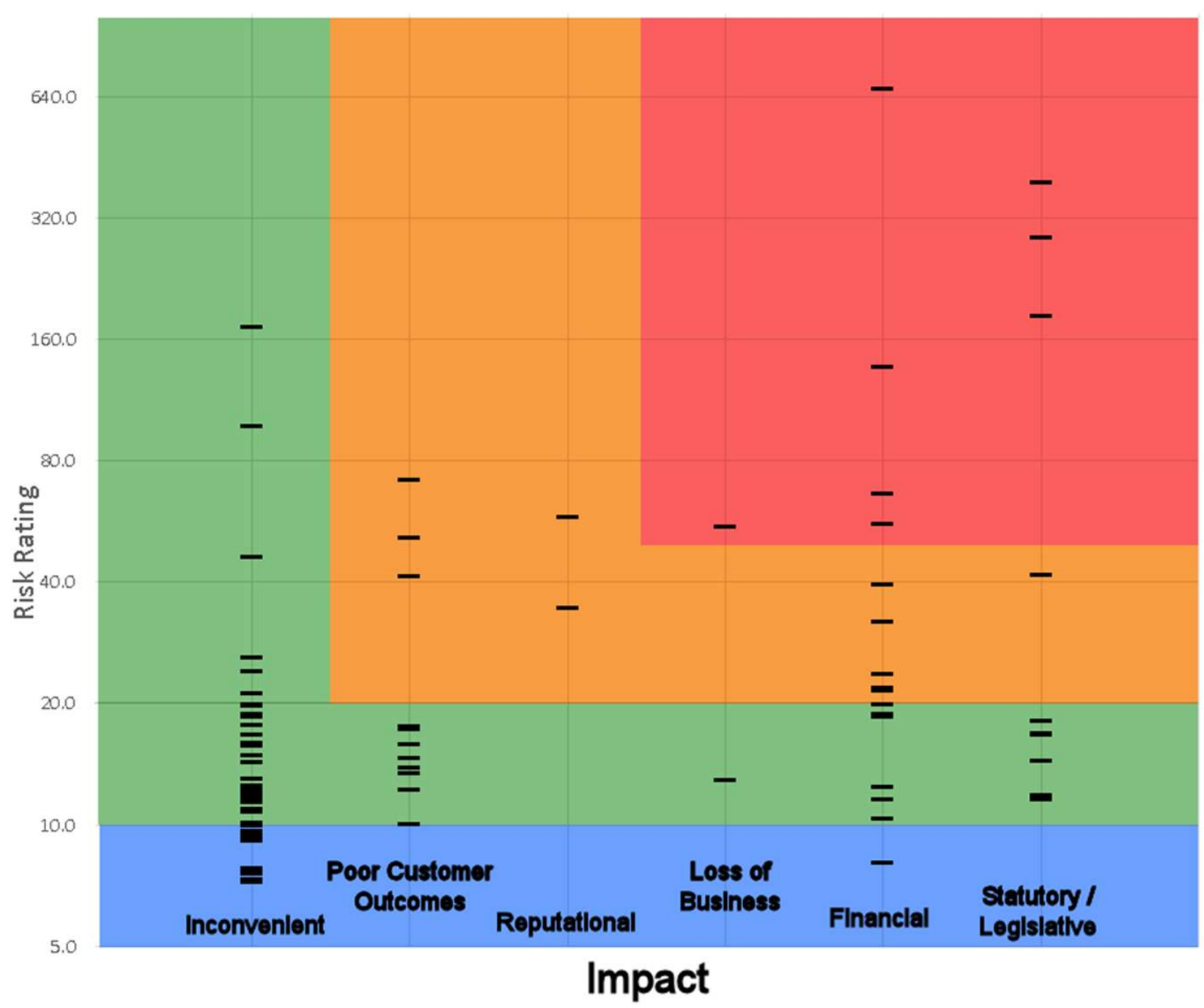

Figure 6: Distribution of EUCAs by Risk Rating and Impact

There were 8 applications in the red quadrant, 14 in the amber, 116 in the green and 20 in the blue. The three highest risks in the paper were recommended to have the most urgent programme for remediation are

(a) An Access database dating back to 2003 with no existing support and carrying substantial value of business. This risk was immediately put on Magique and a replacement system is being sought.

(b) The Complaints Compliance Operations Database, again in Access and without supporting technical documentation. Again, this risk is on Magique and technical documentation is being prepared.

(c) A cluster of spreadsheets used by HR for disciplinary, grievance and absence management exhibiting poor data management in spite of the privacy and access risks being well controlled. This issue, in addition to other business needs, has led to another system being sought to replace these spreadsheets.

The Chief Risk Officer (CRO) approved the paper and this action enabled

- The revised EUC Policy to be included in the Company Controls Documentation.
- Remedial action to be put in each department's business plan for action during 2019 commensurate with the level of risk.

The Executive require an update of the EUC situation in November 2019.

### 6.5 Amendments to the EUC Risk Assessment application

Two amendments were required to the Risk Assessment application and these were discovered whilst preparing for the executive meeting and to streamline the process. They were:

**1. Risk Rating Band**

The Risk Rating Band had been previously calculated from a numeric risk rating which in turn is a function of the materiality, complexity and other controls on the EUC application, but not the impact.

The impact which is collected by the Risk Assessment Application takes one of six values, being

1. Inconvenient
2. Poor Customer Outcomes
3. Reputational
4. Loss of Business
5. Financial
6. Statutory / Legislative

The user who is assessing the application in question provides the highest number out of these six governed by the outcome should the application fail to function correctly.

We realised that if an application was given a red rating the impact upon a risk crystallising would be much less if the impact was classified as inconvenient than it would be if it was (say) financial.

Therefore, we decided to amend the calculation of Risk Rating Band to say that if the impact is "inconvenient" the risk rating band can only be blue or green, and the risk rating band can be red only if the impact is "loss of business", "financial" or "statutory / legislative".

2. **Streamlining the process**

The requirement exists for producing the inventory, KPI reporting and re-assessing each EUCA at least annually and iServer exports metadata about a selection of or all the EUCAs to satisfy this need (see section 4).

### 6.6 Position at the end of 2018

We provided an EUC inventory to all department heads and asked them to keep it up to date in line with the EUC policy.

Figure 7 shows the skew distribution of applications amongst departments, the point being that 85% of the applications are contained within only 7 departments.

Appendix A lists the 7 departments in the left-hand side of Figure 5 and summarises the use of each EUCA. This information is intended to help the reader to find similar EUCAs in his own organisation.

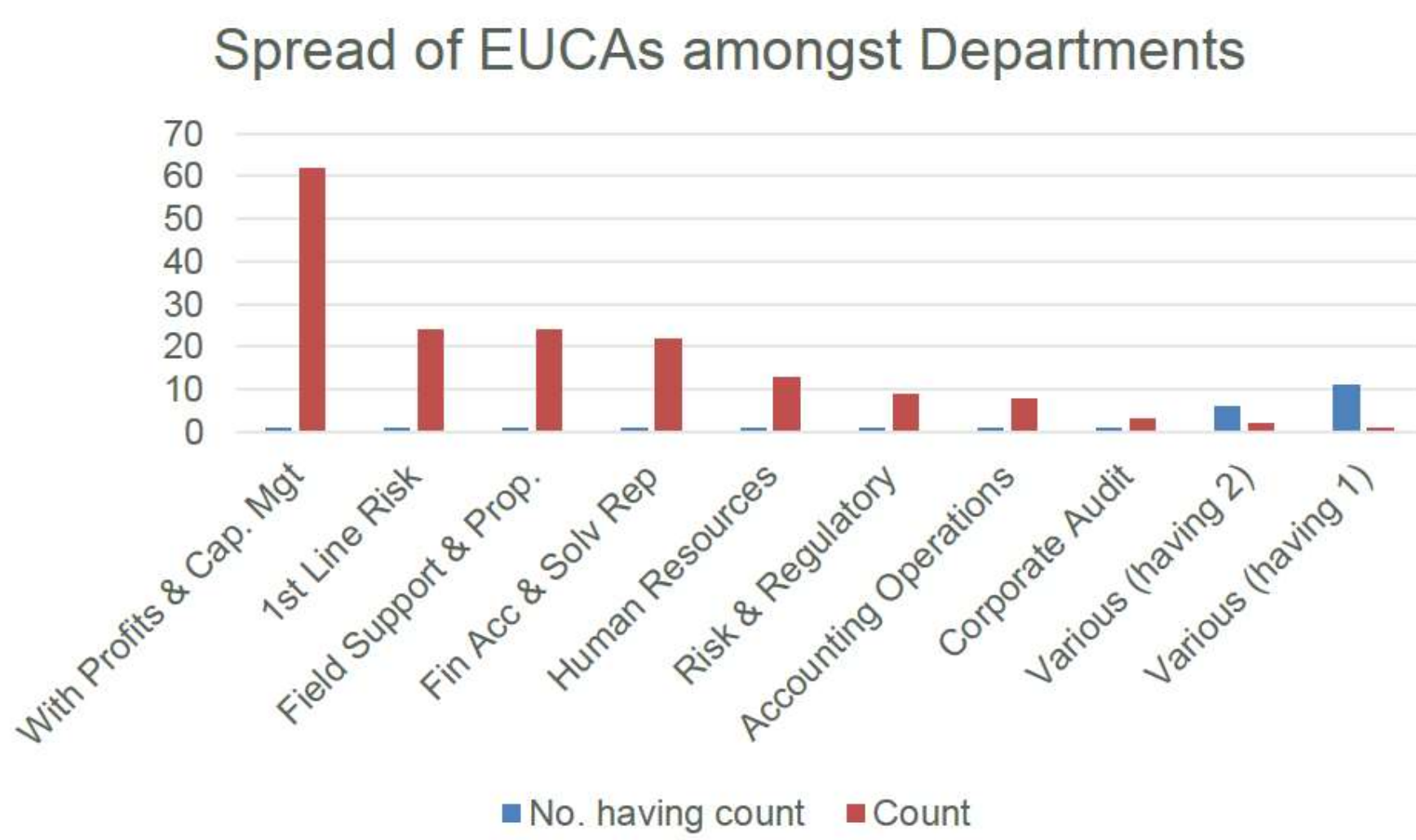

Figure 7: How EUCAs are spread across departments

Those departments with a smaller number of EUCAs (right hand side of Figure 7) could assess these within the time requested. For the remainder we adopted an understanding approach whereby the assessment could be part of the 2019 business plan instead. The plan for these is as follows:

- Accounting Operations: We received a sample of 8 assessments in 2018 and this is the tip of the iceberg. They use thousands of spreadsheets as part of their transaction processing function and one of each kind is up for risk assessment. They are enthusiastic about implementing a phased approach during 2019 and by May the number assessed reached 60.

- Financial Assumptions: This department provided a green rating from the Head Office Managers Meeting the previous April (Section 6.1) so no action is being taken for the time being because we see this as a lower EUC risk. This will be reviewed at some stage in the future.

- With Profits and Capital Management: The manager has already provided information about 62 spreadsheets and knows to submit more assessments as the need arises. No further action needed for the time being.

- Financial Accounting and Solvency Reporting: Spreadsheets of high complexity and materiality were part of one of the pilots in Autumn 2017 and 20 of these revealed a red or amber rating when assessed. Solvency Reporting uses Finsbury Spreadsheet Workbench to provide audit and version control in this area (Finsbury, 2014). Solvency Reporting also successfully uses a Spreadsheet Controls Framework which ensures appropriate peer review whereby results are challenged and locked down with financial and actuarial analysis as necessary. This is achieved by having three tabs on every spreadsheet and these are recognised as positive indicators within the EUC policy. They are:

    - Control: Contains doer & checker evidence and sign-off of the spreadsheet and version history.
    - Validation: Describes the changes, what checks are done, who did them and when, the checker and date and if necessary, the reviewer and date.
    - Documentation: Outlines the purpose of the workbook, details individual sheets, and gives instructions for how to use the spreadsheet.

  The Head of Actuarial rigorously enforces the Spreadsheet Controls Framework on every spreadsheet in the valuation folder by means of a macro to ensure that elements have been completed. These three indicators are in line with the first three items as recommended in the "initial remediation plan" (McGeady & McGouran, 2009, Page 3). Our control is retained within the business and not migrated to a controlled IT environment.

  In addition, audit work is done to ensure that accounts are prepared in line with statutory rules and that regulatory responses are compliant.

  In April 2019 we agreed a plan with the Head of Actuarial to identify any gaps in the Spreadsheet Controls Framework where the EUC Policy is not met and then plan how to amend both the framework and the EUC policy so that the objectives of the EUC policy are still met. To be complete by the end of 2019.

## 7 MITIGATION OF EUC RISK

2018 gave us the opportunity to determine where EUC risk exists within the Wesleyan and how it can be mitigated.

Firstly, why EUC? EUC applications (usually spreadsheets) are written to solve a problem – otherwise why write the spreadsheet? For example, a business need has been satisfied by a system run and supported by IT for years (sometimes decades) and the need is modified or it changes.

Historically a change request takes too long for IT to implement so the department using the system goes its own way and writes a spreadsheet or cluster of spreadsheets to fill the gap. Thereupon the EUC risk arises if the control is lacking and due to any errors in the transmission of data.

The options for areas where there are EUC's are:

(a) To mitigate the risk – in other words to ensure that controls are in place to reduce the likelihood of the risk crystallising. This is a major part of our EUC policy.

(b) To remove the risk – this is to completely avoid or bypass the EUC applications which constitute the risk, sometimes by creating a new system.

(c) To accept the risk. When the risk crystallises, the cost is a better alternative than either of the two options above.

Where EUC applications (especially if they have a red or amber risk rating) cluster in one place we find that putting controls round the EUC applications may not completely solve the problem and the EUC policy encourages and supports option (b) above, for example:

- In HR, errors due to manual transcription of data from one spreadsheet to another, in addition to other business needs, has resulted in a new system for HR and Payroll administration being sought.

- The risks indicated by EUC in Financial Accounting have supported the requirement for an existing project to replace the Finance system and this is now in progress.

- The Access database dating back to 2003 is an example where only option (b) above is appropriate. The fact that it is isolated from our point of sale system means that we are potentially missing out on cross-selling opportunities so a replacement which will integrate with this system adds to the business case. This is also in progress.

## 8 BAU ACTIVITY IN 2019

Activity is ongoing to continually improve the EUC control situation and ensure that a society-wide awareness of the EUC policy continues.

KPI reporting on EUC has been incorporated within Data Governance from March and is supported by monthly requests of each EUC application owning manager in two areas:

### 8.1 Annual review of applications

Each application according to the EUC policy needs to be reviewed annually so each month when one or more applications have come up for review, these are listed and the owning manager is simply asked to confirm that the application is still fit for purpose and in use. The return of this information causes the next review date to be stepped forward a year.

If there are substantial changes to the use, materiality or complexity of the application or if there are any new applications, the owner is asked to download the Risk Assessment Application from the intranet and complete it, returning it to Data Governance.

Particularly in Q1, the returns have indicated that certain applications are no longer in use and were retired and some others indicate a change in ownership. This all helps in keeping the inventory up to date.

### 8.2 Risks entered on the Magique system

iServer keeps a record of whether each red or amber rated application has its risk recorded on the Magique system. If there is no such record on Magique, the monthly communication to the owning manager lists these, asking for the Magique Risk ID and the opening and closing dates on Magique. When remedial action on the application has taken place, the application is re-assessed and the assessment returned to Data Governance.

Where appropriate, risks are recorded on Magique and are tracked by Corporate Audit. This focusses the user's mind on remediating the EUC risk.

We use established methods for turning inherent risk into residual risk and expressing the risk rating as a product of likelihood and severity (Herrera, 2017), (Xenon, 2019).

### 8.3 Monthly Key Performance Indicator (KPI) Reporting

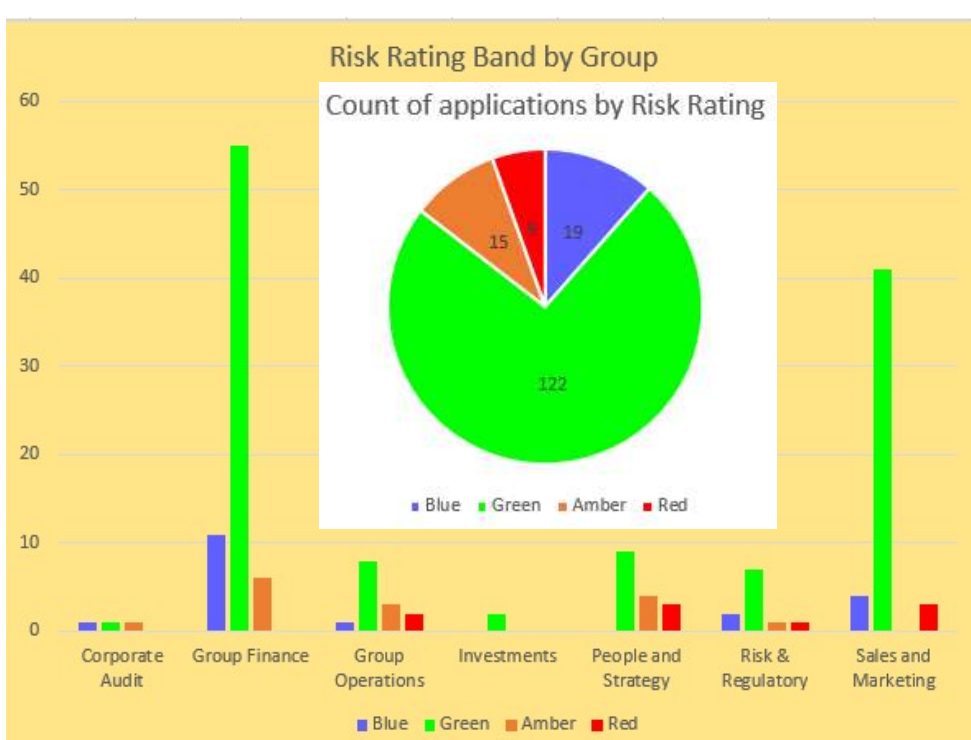

Figure 9: KPI March 2019

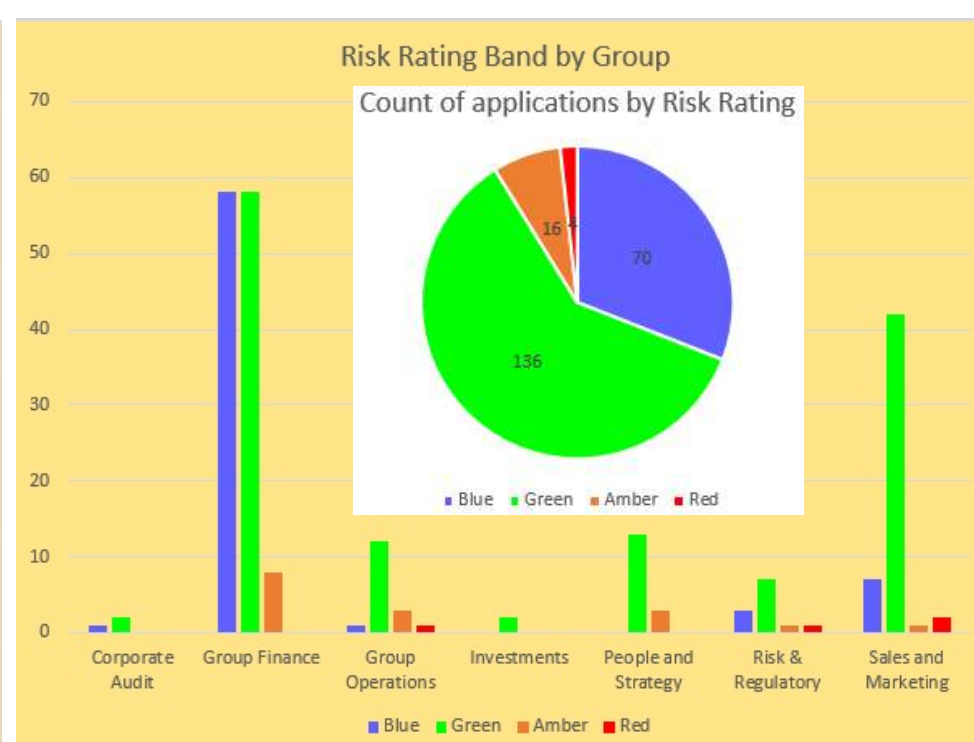

Figure 8: KPI May 2019

Figures 8 and 9 show the progress made between March and May, the main points being:

- Accounting Operations have provided an influx of spreadsheets from the Transaction Processing function, all of which having either a blue or green rating.

- Better control over certain applications having a red rating has been introduced meaning that they now have an amber or green rating.

### 8.4 Reaction from the business to the introduction of the EUC Policy

The business has not been compelled to assess EUC risk for several years prior to the introduction of this policy. Starting the communication at executive level has meant that each department has known in advance that involvement and commitment was expected of them.

Knowledge that a risk-based approach is being used provides an understanding that action needs to be taken to identify and mitigate against, remove or accept the most serious risks and this has certainly helped.

Reaction is mainly as follows:

- There is general acceptance of the policy because of the way in which it was introduced.

- 108 (81%) (see figure 4) indicated that there is negligible risk so no action is being taken for the time being (or unless they get caught out)! When this happens, they are only too keen to comply because they are made aware of the risks.

- The remainder had action to do, nearly everyone accepted and are using the policy (see section 6.6). Some departments are quite enthusiastic in wanting to comply.

- Where there are existing controls (Actuarial) there was quite understandably resistance from the point of view of having to change and because of the additional work. We are working together to achieve the objectives of the EUC policy.

- Resistance is at its greatest where the policy demands that an EUC deficiency calls for an entry to be made on Magique (the risk register) because this exposes the risk to Internal Audit and eventually gets executive attention. Wherever possible we encourage the department to introduce appropriate EUC controls which will avoid an entry being made on Magique.

One of the most visible issues is the matter of communication. The policy is well documented on the intranet and offers to help by showing what to do in various situations is always accepted.

## 9 INTERESTING SURPRISES

It's amazing what you find if you look (or what you miss if you don't)! When we visited one department, we were expecting to receive resistance to the governance which EUC controls would apply. Instead we found that the manager was only too keen to cooperate because this policy has allowed her to justify key changes to her department's system. The advent of the EUC policy has brought the matter to executive attention and a plan is now in place to replace the system.

We adopted a top-down approach for identification which means that each manager recognises his or her areas at risk. We were 'tipped off' concerning the administration of one system. We discovered that it is administered using a cluster of spreadsheets which overcome shortcomings in some very old IT systems. We saw the manager the same day only to find that the sole person who could look after the spreadsheets was leaving within a month. Somebody else was quickly brought in to be trained up to sufficient level to mitigate the EUC risks.

## 10 THE EUC RISK ASSESSMENT APPLICATION

The two main screens are given here.

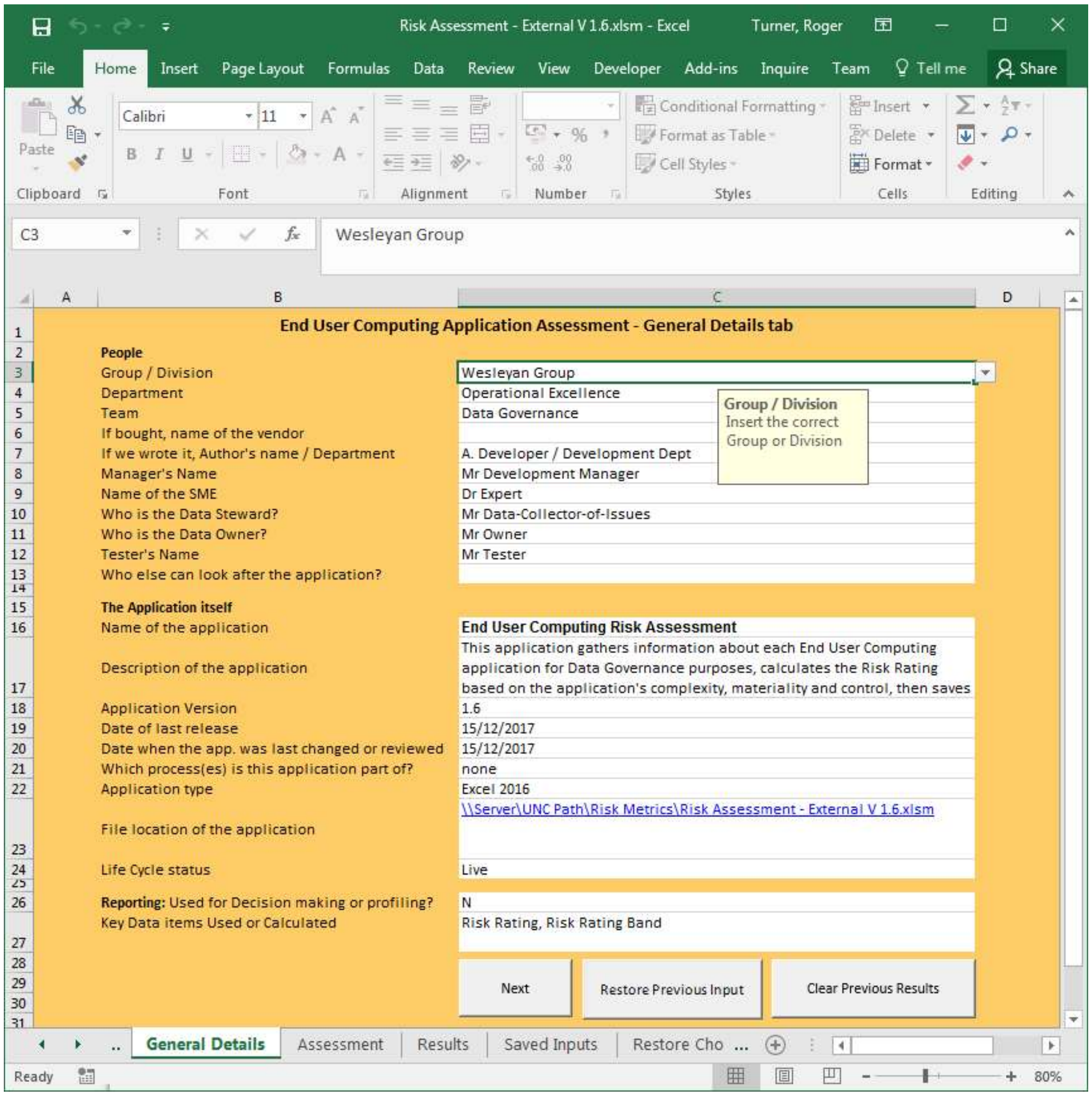

Figure 10- Risk Assessment Application screen - General Details

The top part of the screen is all about the people who interact with the application. We in Data Governance use these people as a point of contact.

The next section is about the application itself, giving its name, description, version and version history, where it is and which platform it runs on. This provides us with more depth to the information we hold about the application. All of these details are recorded in iServer.

We learned from one of the pilots that users like to partly complete a batch of applications and go back to them later to finish off. We needed to provide a “Restore Previous Input” button which allows the user to call back information about any previously entered application for completion.

Clicking on “Next” navigates us to the next screen.

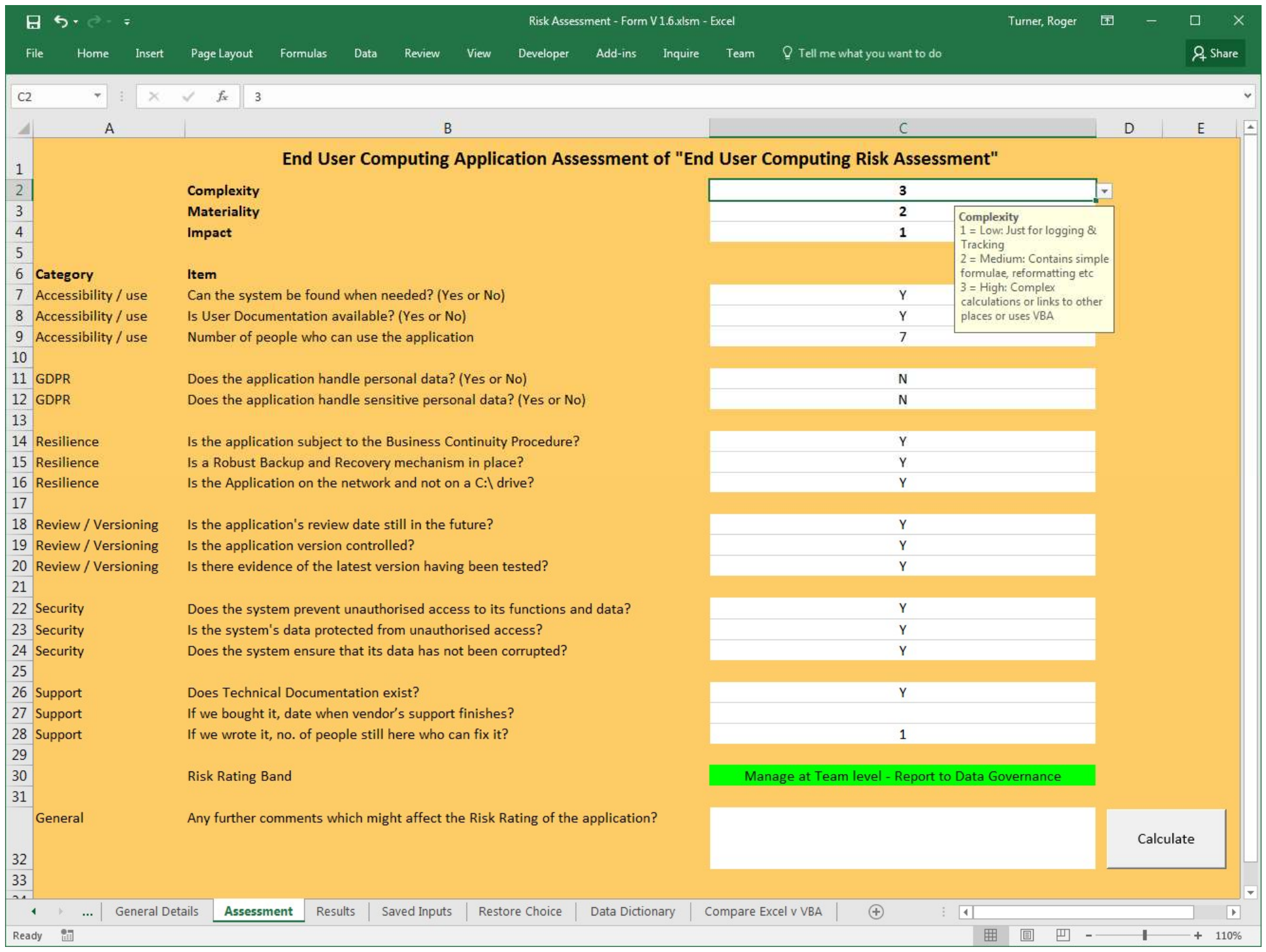

Figure 11- Risk Assessment Application screen - Assessment

The second screen gathers information about the control of the application. The SME or other expert in the User Department enters the Complexity and Materiality ratings first and then answers mostly "yes" or "no" to the other questions, divided into the categories as described above.

The last (General) box is to include any free-form information about the application which might be useful, for example "We are currently training a second person who can fix this application if it breaks".

At the end of entering all the data, the user clicks on the "Calculate" button and the Risk Assessment application calculates the risk rating band according to the details entered.

The green box near the bottom is then populated with the next action required, according to whether the rating is blue, green, amber or red. It is coloured appropriately. The possible outcomes are:

a. Blue – No action needed.

b. Green – An awareness of this application is needed. This is the minimum rating band for applications which hold customer data (GDPR) and applications which are Green and above are reported to Data Governance and subject to annual review.

c. Amber – Falls short of acceptable control and an action plan is needed to fix. An entry is made in the Magique system. We would expect the plan to be implemented within three months.

d. Red – As Amber but urgent action is needed to fix – within a month or before the application is next run if later.

The last box is for free-format information about the risk which might assist in mitigation.

The Risk Assessment application calculates the next review date as being a year from the previous review and the applications which carry the most significant degree of risk are known.

We updated the EUC policy having run the pilots.

Already some areas see the policy favourably. In rolling the policy out a risk-based approach will be used so that areas where there is a greater risk will be worked with first.

## 11 CONCLUSION

Wesleyan started with an EUC policy which required updating. After we had written a new policy and the EUC Risk Assessment Application we successfully ran pilots in two areas of Finance. The assessment results from these pilots enabled some quick wins to be done and from the learning points gained we were able to improve the policy and the Risk Assessment Application.

We have worked successfully during the last year to increase the exposure and awareness of EUC to the Wesleyan. We have used a top-down approach to identify areas of greatest risk. Buy-in at Executive level from the start of the year was essential because only by this means can meaningful resource be devoted to areas of need.

A flexible and understanding approach has yielded dividends. EUC control generally takes a back seat in relation to business priorities and if an action to remediate an application cannot be completed immediately because of lack of resource then we have been able to agree a plan of action.

KPI reporting has started and the Executive are looking to see an improvement in the EUC estate by the end of 2019. At the time of writing this is already evident. Now that the momentum is established, it is important not to let it go to waste in subsequent months or years.

Any models or information contained in this paper are intended for educational purposes only. To the extent permitted by law, the author and Wesleyan Assurance Society shall not be held liable for any liability or loss suffered by a third party who uses the models or information within this document for purposes for which they were not intended.

## REFERENCES

**Bregar**, (2004), 'Complexity Metrics for Spreadsheet Models', Online [available] https://arxiv.org/ftp/arxiv/papers/0802/0802.3895.pdf, accessed 13/11/2020

**Chambers & Hamill**, (2008), 'Controlling EUC Applications – a case study', Online [available] https://arxiv.org/ftp/arxiv/papers/0809/0809.3595.pdf, accessed 13/11/2020

**Chartis,** (2016), 'Quantification of EUC Risk in Financial Services', Online [available], http://www.clusterseven.com/wp-content/uploads/2016/07/Quantification-of-EUC-Risk-Final.pdf , accessed 13/11/2020. Source: Chartis Research, Quantification of EUC Risk in Financial Services, June 2016.

**CIR Magazine**, (2018), 'Shortlist 2018', Online [available] http://www.cirmagazine.com/riskmanagementawards/shortlist18.php, accessed 13/11/2020

**The Corporate IT Forum**, (2016), 'Generating Business Benefit from Shadow IT', Online [available] https://www.corporateitforum.com/event/workshop/1216-citizen-itshadow-it, accessed 28/03/2018. CITF aims to provide effective platforms for discussion and exchange of information between technology peers, and aims to provide a base for developing common views and processes on business and technology issues.

**EuSpRIG**, (2018), Horror Stories, Online [available] http://www.eusprig.org/horror-stories.htm accessed 13/11/2020

**Financial Times**, (2013), January 21st, 2013 edition of the Financial Times, Online [available] https://ftalphaville.ft.com/2013/01/21/1344742/can-haz-spredshetz/ , accessed 13/11/2020 by putting 'can-haz-spredshetz' into Google.

**Finsbury**, (2014), 'EUC Enterprise', Online [available], http://finsburysolutions.com/products-overview/, accessed 13/11/2020

**George Mallikourtis CISA, CISM & Efthimis Papanikolaou**, CISA, ISMS IA, Hellenic American Union Conference, (2010), 'EUC (EUC) Risk: From Assessment to Audit', Online [available] http://conferences.hau.gr/resources/aifs2010/proceedings10/mallikourtispapanikolaou-2.pdf , accessed 13/11/2020

**Herrera**, (2017), 'What is Residual Risk (& How Do You Calculate It)?', Online [available] https://bcmmetrics.com/what-is-residual-risk-and-how-to-calculate-it/, accessed 13/11/2020

**IT Governance**, (2018), 'The EU General Data Protection Regulation (GDPR)', Online [available], https://www.itgovernance.co.uk/data-protection-dpa-and-eu-data-protection-regulation, accessed 13/11/2020

**Magique Galileo Risk Management**, (2018), Online [available] http://magiquegalileo.com/ , accessed 13/11/2020. Magique is a Risk Management system that covers the recording, assessment and approval of risks, controls and events. The system also covers monitoring, trend analysis and reporting. Hosted on a Microsoft Windows 2012 Server.

**McGeady & McGouran**, (2009), 'EUC in AIB Capital Markets: A Management Summary', Online [available] https://pdfs.semanticscholar.org/189b/e31ecc2cedc26c78a6d3818ba9cdcd564203.pdf, accessed 13/11/2020

**Microsoft**, (2013), 'Discovery and Risk Assessment Server 2013', Online [available], https://technet.microsoft.com/en-us/library/jj612849.aspx, accessed 13/11/2020

**Microsoft,** (2016), 'Spreadsheet Inquire in Excel 2016 for Windows', Online [available] https://support.office.com/en-gb/article/what-you-can-do-with-spreadsheet-inquire-in-excel-2016-for-windows-5444eb12-14a2-4d82-b527-45b9884f98cf, accessed 13/11/2020

**Microsoft** (2018), 'Business Intelligence', Online [available] https://powerbi.microsoft.com/en-us/, accessed 28/03/2018. Power BI is provided by Microsoft and has a desktop application which allows you to connect to data from a multiple of sources, to shape that data through queries and use the results to create reports using a range of standard and bespoke visuals. The resultant report files can be shared like any other or can be uploaded (and shared) on the Power BI Service which is cloud based.

**Orbus Software**, (2018), Capabilities, Online [available] https://www.orbussoftware.com/business-architecture/business-capability-view, accessed 13/11/2020

**PwC,** (2004), 'The Use of Spreadsheets: Considerations for section 404 of the Sarbanes-Oxley Act', Online [available] http://www.spreadsheetdetective.com/main/PwC-SpreadsheetsSoX.pdf, accessed 13/11/2020

**Wikipedia,** (2018), 'Microsoft Excel – Early History', Online [available] https://en.wikipedia.org/wiki/Microsoft_Excel#Early_history, accessed 13/11/2020

**Xenon Group**, (2019), 'The Risk Formula – How to calculate the level of risk to your business', Online [available] http://www.xenongroup.co.uk/knowledge-centre/risk-management/the-risk-formula-how-to-calculate-the-level-of-risk-to-your-business, accessed 13/11/2020

## APPENDIX A – USE OF SPREADSHEETS IN BUSINESS AREAS

This appendix is a brief summary of the use to which spreadsheets are put, in the areas which use the most spreadsheets.

- With Profits and Capital Management, and Solvency Monitoring – Actuarial – keeps the Society's financial position up to date and provides information to support Solvency II legislation.

- Financial Accounting and Accounting Operations – Accounting – Preparation of the Wesleyan's accounts, receipts and payments.

- Field Support and Proposition and 1st Line Risk – Support for the Financial Consultants and logging of brokered business

- Human Resources – Joiners, Movers, Leavers, Benchmarking, employee relationship activity, workflow

- Risk & Regulatory – Work management, Regulatory changes

## APPENDIX B – ASSESSMENT TEMPLATE

Screenshot of the Assessment Template provided to those attending the Head Office Managers Meeting, April 2018 (See section 6.1)

**End User Computing Scoring template**

**Purpose: To assist line managers with assessing the level of risk associated with use of spreadsheets and other applications as applicable.**
**This will take no more than 10 minutes to complete.**
**Use the Tab button to navigate from one field to the next.**

| | |
|---|---|
| Your name | **A Manager** |
| Name of your Department | **Any Department** |

**Have you any spreadsheets or applications which are not supported by any of the Wesleyan Group's IT departments and which are in any way material to the Business? Enter 1 for Yes, 0 for No**

| | |
|---|---|
| **If you have, put a 1 here and continue with the rest of this form.**<br>**If you haven't, put a 0 here and go straight to the end to read the instruction.** | **1** |

**Pick the process which has the most complex calculations, or which has the most material impact on the Business operation**

| | |
|---|---|
| **Name of the process, and in which business area** | Complex Calculation Process |
| **Materiality of the process: Insert 1 to 3 as appropriate.**<br>1 = Internal operations or day to day decisions only<br>2 = Management Reporting or making key business decisions<br>3 = Financial or Regulatory Reporting or Private / Confidential information | **2** |
| **Complexity of the spreadsheet or application in the process: Insert 1 to 3 as appropriate**<br>1 = No calculations, just logging and information tracking<br>2 = Simple formulae or reformatting data<br>3 = Complex formulae, external references, any programming language | **3** |

**Control: Score 1 (bad) to 3 (good) for each of these:**
(decimals, for example 1.5, 2.5 are allowed)

| | |
|---|---|
| How confident are you that there is the knowledge in the department able to fix an application if it breaks after the person who wrote it has left? | **1.5** |
| Does this area handle personal data or sensitive personal data (GDPR)? Put 0 for no, 1 for yes | **0** |
| How confident are you the processes will continue get good, reliable results even if you're short staffed in key areas? | **2** |
| If there is loss of service, equipment breaks or if there is a power outage how confident are you that you can get back to a working state quickly? | **3** |
| Have you confidence that controls are in place to ensure that the latest version is being used when needed | **2** |
| How confident are you that the applications don't give misleading results resulting from misuse or unauthorised access (which might be due to malicious intent)? | **3** |

**Please review your responsibilities regarding End User Computing as described within the policy document, link given below**

**You are Amber. Action is needed. Return this spreadsheet to Data Governance. Your spreadsheets and applications need to be assessed, errant ones recorded on Magique and there needs to be an action plan to fix.**

Figure 12: Departmental EUC Risk Assessment Template

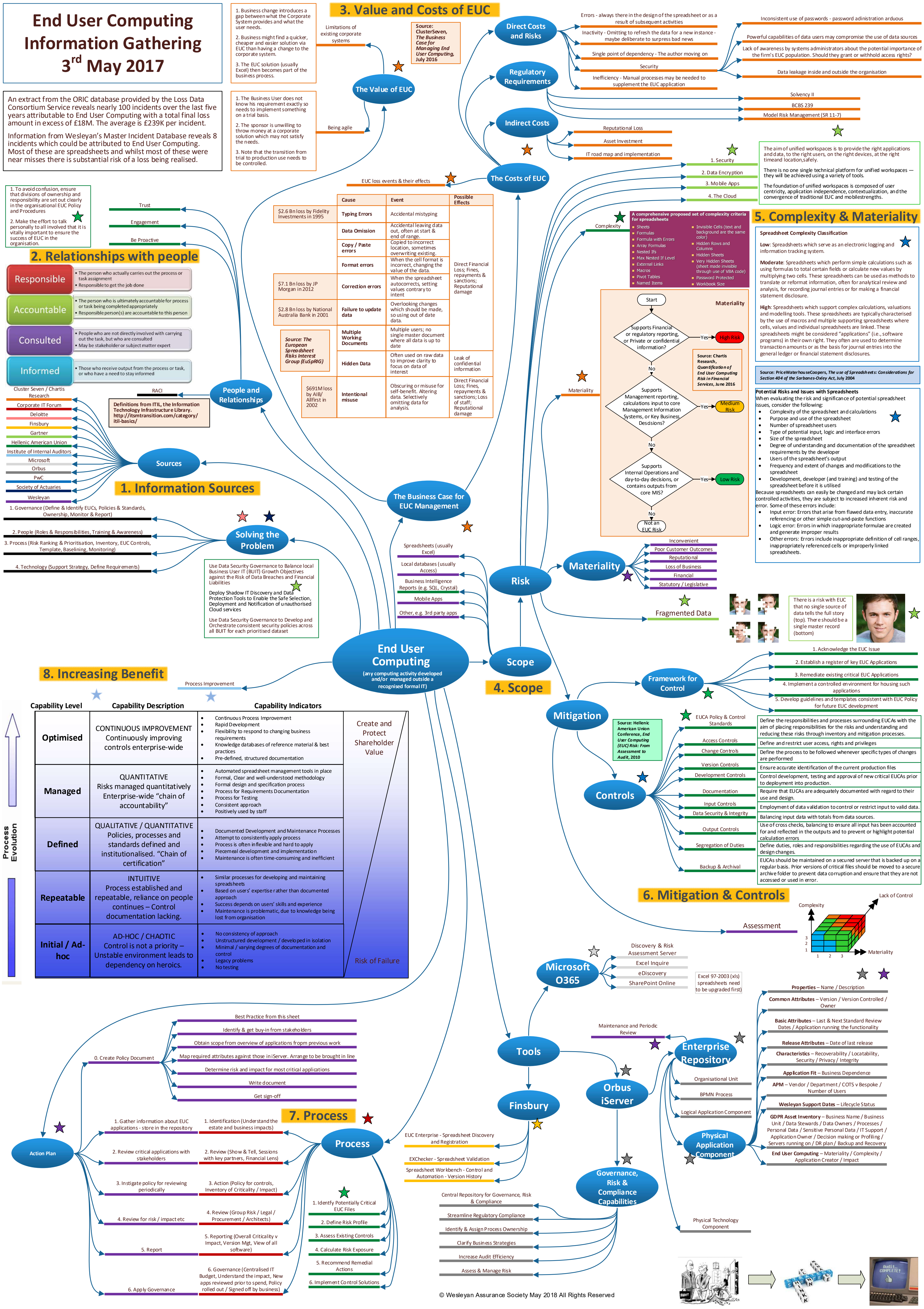
End User Computing Information Gathering 3rd May 2017
An extract from the ORIC database provided by the Loss Data Consortium Service reveals nearly 100 incidents over the last five years attributable to End User Computing with a total final loss amount in excess of £18M. The average is £239K per incident.
Information from Wesleyan's Master Incident Database reveals 8 incidents which could be attributed to End User Computing. Most of these are spreadsheets and whilst most of these were near misses there is substantial risk of a loss being realised.
1. Business change introduces a gap between what the Corporate System provides and what the user needs.
2. Business might find a quicker, cheaper and easier solution via EUC than having a change to the corporate system.
3. The EUC solution (usually Excel) then becomes part of the business process.
Limitations of existing corporate systems
1. The Business User does not know his requirement exactly so needs to implement something on a trial basis.
2. The sponsor is unwilling to throw money at a corporate solution which may not satisfy the needs.
3. Note that the transition from trial to production use needs to be controlled.
Being agile
3. Value and Costs of EUC
Source: ClusterSeven, The Business Case for Managing End User Computing, July 2016
The Value of EUC
Direct Costs and Risks
Regulatory Requirements
Indirect Costs
The Costs of EUC
Errors - always there in the design of the spreadsheet or as a result of subsequent activities
Inactivity - Omitting to refresh the data for a new instance - maybe deliberate to surpress bad news
Single point of dependency - The author moving on
Security
Inefficiency - Manual processes may be needed to supplement the EUC application
Inconsistent use of passwords - password adminstration arduous
Powerful capabilities of data users may compromise the use of data sources
Lack of awareness by systems administrators about the potential importance of the firm's EUC population. Should they grant or withhold access rights?
Data leakage inside and outside the organisation
Solvency II
BCBS 239
Model Risk Management (SR 11-7)
Reputational Loss
Asset Investment
IT road map and implementation
1. Security
2. Data Encryption
3. Mobile Apps
4. The Cloud
The aim of unified workspaces is to provide the right applications and data, to the right users, on the right devices, at the right time and location, safely.
There is no one single technical platform for unified workspaces — they will be achieved using a variety of tools.
The foundation of unified workspaces is composed of user centricity, application independence, contextualization, and the convergence of traditional EUC and mobile strengths.
EUC loss events & their effects
Cause | Event | Possible Effects
Typing Errors | Accidental mistyping
Data Omission | Accidental leaving data out, often at start & end of range.
Copy / Paste errors | Copied to incorrect location, sometimes overwriting existing.
Format errors | When the cell format is incorrect, changing the value of the data.
Correction errors | When the spreadsheet autocorrects, setting values contrary to intent
Failure to update data | Overlooking changes which should be made, so using out of date data.
Multiple Working Documents | Multiple users; no single master document where all data is up to date
Hidden Data | Often used on raw data to improve clarity to focus on data of interest
Intentional misuse | Obscuring or misuse for self-benefit. Altering data. Selectively omitting data for analysis.
Direct Financial Loss; Fines, repayments & sanctions; Reputational damage
Leak of confidential information
Direct Financial Loss; Fines, repayments & sanctions; Loss of staff; Reputational damage
$2.6 Bn loss by Fidelity Investments in 1995
$7.1 Bn loss by JP Morgan in 2012
$2.8 Bn loss by National Australia Bank in 2001
Source: The European Spreadsheet Risks Interest Group (EuSpRIG)
$691M loss by AIB/ Allfirst in 2002
1. To avoid confusion, ensure that divisions of ownership and responsibility are set out clearly in the organisational EUC Policy and Procedures
2. Make the effort to talk personally to all involved that it is vitally important to the success of EUC in the organisation.
Trust
Engagement
Be Proactive
2. Relationships with people
Responsible
The person who actually carries out the process or task assignment
Responsible to get the job done
Accountable
The person who is ultimately accountable for process or task being completed appropriately
Responsible person(s) are accountable to this person
Consulted
People who are not directly involved with carrying out the task, but who are consulted
May be stakeholder or subject matter expert
Informed
Those who receive output from the process or task, or who have a need to stay informed
People and Relationships
RACI
Definitions from ITIL, the Information Technology Infrastructure Library. http://itsmtransition.com/category/itil-basics/
Cluster Seven / Chartis Research
Corporate IT Forum
Deloitte
Finsbury
Gartner
Hellenic American Union
Institute of Internal Auditors
Microsoft
Orbus
PwC
Society of Actuaries
Wesleyan
Sources
1. Information Sources
1. Governance (Define & Identify EUCs, Policies & Standards, Ownership, Monitor & Report)
2. People (Roles & Responsibilities, Training & Awareness)
3. Process (Risk Ranking & Prioritisation, Inventory, EUC Controls, Template, Baselining, Monitoring)
4. Technology (Support Strategy, Define Requirements)
Solving the Problem
Use Data Security Governance to Balance local Business User IT (BUIT) Growth Objectives against the Risk of Data Breaches and Financial Liabilities
Deploy Shadow IT Discovery and Data Protection Tools to Enable the Safe Selection, Deployment and Notification of unauthorised Cloud services
Use Data Security Governance to Develop and Orchestrate consistent security policies across all BUIT for each prioritised dataset
The Business Case for EUC Management
5. Complexity & Materiality
Complexity
A comprehensive proposed set of complexity criteria for spreadsheets
Sheets, Formulas, Formula with Errors, Array Formulas, Nested Ifs, Max Nested IF Level, External Links, Macros, Pivot Tables, Named Items
Invisible Cells (text and background are the same color), Hidden Rows and Columns, Hidden Sheets, Very Hidden Sheets (sheet made invisible through use of VBA code), Password Protected, Workbook Size
Spreadsheet Complexity Classification
Low: Spreadsheets which serve as an electronic logging and information tracking system.
Moderate: Spreadsheets which perform simple calculations such as using formulas to total certain fields or calculate new values by multiplying two cells. These spreadsheets can be used as methods to translate or reformat information, often for analytical review and analysis, for recording journal entries or for making a financial statement disclosure.
High: Spreadsheets which support complex calculations, valuations and modelling tools. These spreadsheets are typically characterised by the use of macros and multiple supporting spreadsheets where cells, values and individual spreadsheets are linked. These spreadsheets might be considered "applications" (i.e., software programs) in their own right. They often are used to determine transaction amounts or as the basis for journal entries into the general ledger or financial statement disclosures.
Source: PriceWaterhouseCoopers, The use of Spreadsheets: Considerations for Section 404 of the Sarbanes-Oxley Act, July 2004
Materiality
Start
Supports Financial or regulatory reporting, or Private or confidential information?
Yes → High Risk
No
Supports Management reporting, calculations input to core Management Information Systems, or Key Business Decisions?
Yes → Medium Risk
Supports Internal Operations and day-to-day decisions, or contains outputs from core MIS?
Yes → Low Risk
Not an EUC Risk
Source: Chartis Research, Quantification of End User Computing Risk in Financial Services, June 2016
Potential Risks and Issues with Spreadsheets
When evaluating the risk and significance of potential spreadsheet issues, consider the following:
Complexity of the spreadsheet and calculations
Purpose and use of the spreadsheet
Number of spreadsheet users
Type of potential input, logic and interface errors
Size of the spreadsheet
Degree of understanding and documentation of the spreadsheet requirements by the developer
Users of the spreadsheet's output
Frequency and extent of changes and modifications to the spreadsheet
Development, developer (and training) and testing of the spreadsheet before it is utilised
Because spreadsheets can easily be changed and may lack certain controlled activities, they are subject to increased inherent risk and error. Some of these errors include:
Input error: Errors that arise from flawed data entry, inaccurate referencing or other simple cut-and-paste functions
Logic error: Errors in which inappropriate formulae are created and generate improper results
Other errors: Errors include inappropriate definition of cell ranges, inappropriately referenced cells or improperly linked spreadsheets.
Spreadsheets (usually Excel)
Local databases (usually Access)
Business Intelligence Reports (e.g. SQL, Crystal)
Mobile Apps
Other, e.g. 3rd party apps
Risk
Materiality
Inconvenient
Poor Customer Outcomes
Reputational
Loss of Business
Financial
Statutory / Legislative
Fragmented Data
There is a risk with EUC that no single source of data tells the full story (top). There should be a single master record (bottom)
End User Computing
(any computing activity developed and/or managed outside a recognised formal IT)
Scope
4. Scope
Framework for Control
1. Acknowledge the EUC Issue
2. Establish a register of key EUC Applications
3. Remediate existing critical EUC Applications
4. Implement a controlled environment for housing such applications
5. Develop guidelines and templates consistent with EUC Policy for future EUC development
Mitigation
Source: Hellenic American Union Conference, End User Computing (EUC) Risk: From Assessment to Audit, 2010
Controls
EUCA Policy & Control Standards
Define the responsibilities and processes surrounding EUCAs with the aim of placing responsibilities for the risks and understanding and reducing these risks through inventory and mitigation processes.
Access Controls
Define and restrict user access, rights and privileges
Change Controls
Define the process to be followed whenever specific types of changes are performed
Version Controls
Ensure accurate identification of the current production files
Development Controls
Control development, testing and approval of new critical EUCAs prior to deployment into production.
Documentation
Require that EUCAs are adequately documented with regard to their use and design.
Input Controls
Employment of data validation to control or restrict input to valid data.
Data Security & Integrity
Balancing input data with totals from data sources.
Output Controls
Use of cross checks, balancing to ensure all input has been accounted for and reflected in the outputs and to prevent or highlight potential calculation errors
Segregation of Duties
Define duties, roles and responsibilities regarding the use of EUCAs and design changes.
Backup & Archival
EUCAs should be maintained on a secured server that is backed up on a regular basis. Prior versions of critical files should be moved to a secure archive folder to prevent data corruption and ensure that they are not accessed or used in error.
6. Mitigation & Controls
Assessment
Complexity
Lack of Control
Materiality
8. Increasing Benefit
Process Improvement
Capability Level | Capability Description | Capability Indicators
Optimised | CONTINUOUS IMPROVEMENT Continuously improving controls enterprise-wide | Continuous Process Improvement; Rapid Development; Flexibility to respond to changing business requirements; Knowledge databases of reference material & best practices; Pre-defined, structured documentation
Managed | QUANTITATIVE Risks managed quantitatively Enterprise-wide "chain of accountability" | Automated spreadsheet management tools in place; Formal, Clear and well-understood methodology; Formal design and specification process; Process for Requirements Documentation; Process for Testing; Consistent approach; Positively used by staff
Defined | QUALITATIVE / QUANTITATIVE Policies, processes and standards defined and institutionalised. "Chain of certification" | Documented Development and Maintenance Processes; Attempt to consistently apply process; Process is often inflexible and hard to apply; Piecemeal development and implementation; Maintenance is often time-consuming and inefficient
Repeatable | INTUITIVE Process established and repeatable, reliance on people continues – Control documentation lacking. | Similar processes for developing and maintaining spreadsheets; Based on users' expertise rather than documented approach; Success depends on users' skills and experience; Maintenance is problematic, due to knowledge being lost from organisation
Initial / Ad-hoc | AD-HOC / CHAOTIC Control is not a priority – Unstable environment leads to dependency on heroics. | No consistency of approach; Unstructured development / developed in isolation; Minimal / varying degrees of documentation and control; Legacy problems; No testing
Process Evolution
Create and Protect Shareholder Value
Risk of Failure
Microsoft O365
Discovery & Risk Assessment Server
Excel Inquire
eDiscovery
SharePoint Online
Excel 97-2003 (xls) spreadsheets need to be upgraded first)
Tools
Finsbury
Orbus iServer
Enterprise Repository
Maintenance and Periodic Review
Organisational Unit
BPMN Process
Logical Application Component
Physical Application Component
Physical Technology Component
Governance, Risk & Compliance Capabilities
Properties – Name / Description
Common Attributes – Version / Version Controlled / Owner
Basic Attributes – Last & Next Standard Review Dates / Application running the functionality
Release Attributes – Date of last release
Characteristics – Recoverability / Locatability, Security / Privacy / Integrity
Application Fit – Business Dependence
APM – Vendor / Department / COTS v Bespoke / Number of Users
Wesleyan Support Dates – Lifecycle Status
GDPR Asset Inventory – Business Name / Business Unit / Data Stewards / Data Owners / Processes / Personal Data / Sensitive Personal Data / IT Support / Application Owner / Decision making or Profiling / Servers running on / DR plan / Backup and Recovery
End User Computing – Materiality / Complexity / Application Creator / Impact
EUC Enterprise - Spreadsheet Discovery and Registration
EXChecker - Spreadsheet Validation
Spreadsheet Workbench - Control and Automation - Version History
Central Repository for Governance, Risk & Compliance
Streamline Regulatory Compliance
Identify & Assign Process Ownership
Clarify Business Strategies
Increase Audit Efficiency
Assess & Manage Risk
Best Practice from this sheet
Identify & get buy-in from stakeholders
Obtain scope from overview of applications from previous work
Map required attributes against those in iServer. Arrange to be brought in line
Determine risk and impact for most critical applications
Write document
Get sign-off
0. Create Policy Document
7. Process
Process
Action Plan
1. Gather information about EUC applications - store in the repository
2. Review critical applications with stakeholders
3. Instigate policy for reviewing periodically
4. Review for risk / impact etc
5. Report
6. Apply Governance
1. Identification (Understand the estate and business impacts)
2. Review (Show & Tell, Sessions with key partners, Financial Lens)
3. Action (Policy for controls, Inventory of Criticality / Impact)
4. Review (Group Risk / Legal / Procurement / Architects)
5. Reporting (Overall Criticality v Impact, Version Mgt, View of all software)
6. Governance (Centralised IT Budget, Understand the impact, New apps reviewed prior to spend, Policy rolled out / Signed off by business)
1. Identify Potentially Critical EUC Files
2. Define Risk Profile
3. Assess Existing Controls
4. Calculate Risk Exposure
5. Recommend Remedial Actions
6. Implement Control Solutions
© Wesleyan Assurance Society May 2018 All Rights Reserved